\documentclass{appolb}
\usepackage{epsfig}

\begin{document}

\title{Local invariants in effective hydrodynamics of trapped
dilute-gas Bose--Einstein condensates}
\author{Alexander V. Zhukov
\address{Institute of Physics, Academia Sinica, Nankang,
Taipei 11529, Taiwan} }
\maketitle

\begin{abstract}
In the framework of mean-field approximation the dynamics of
Bose-Einstein condensates can be described by the
hydrodynamic-like equations. These equations are analyzed here
with account of mutual interaction between condensate and
non-condensate atoms. The Lagrange invariants and freezing-in
invariants of such a system have been found. This allows to get
some necessary conditions for creation of an atom laser with
controlled parameters of the beam. Particularly, the atom laser
beam can carry quite well defined angular momentum. This can be
practically realized in the most simple case, when the vorticity
of condensate appears to be a freezing-in field. The optimal
conditions for a source mode regime are found out in the paper.
\end{abstract}

\section{Introduction}

One of the youngest, but very rapidly developing field of modern
low temperature physics is the problem of trapped Bose gases at
very low temperatures, when Bose-Einstein condensation occurs. For
recent reviews of this topic in general see Refs.
\cite{1,2,3,4}. In last few years a considerable progress
has been achieved in understanding of the dynamics of
Bose-Einstein condensates (BEC) both at $T=0$ \cite{1} and at
finite temperatures \cite{2,5,10}. However, there many
questions remain in the latter case yet (see discussion in
\cite{11}). To realize the full potential of recent developments
in BEC physics, and to analyze adequately the data of the
experiments in a quantitatively meaningful manner, a detailed
understanding of the BEC dynamics in all its aspects is required.

\par
A detailed investigation of the dynamics (and, particularly,
collective processes) of BEC at various temperatures is extremely
important in view of the possible applications of unique features
of such systems. In all likelihood the most intriguing
experimental project associated with trapped atomic gases is the
so-called `atom laser' or, in other words, highly coherent atomic
beam generator \cite{12,13} (for a review see also
\cite{4}). The proposed experimental configurations should
satisfy a number of basic criteria in order to be called an atom
laser. Of course, the high phase coherence of atomic beam is
required first of all. However, the highly dispersive nature of
the BEC suggests that the spatial focusing and stability of cross
section of a BEC beam will present possibly more of a problem than
encountered in the process of focusing laser light \cite{14}.
Very recently the quasi-continuous atom laser has been constructed
\cite{15}. Furthermore the same group demonstrated the
successful atom optical manipulation, such as reflection, focusing
and the beam storage in a resonator \cite{16} (however, see the earlier
successful works \cite{41,42} based on rather different principle). Thus, another
interesting task appears, namely how to create the atom laser with
well controlled characteristics. Solution of such a problem can
bring atom laser closer to be a useful tool in various potential
applications. The present paper solves one of the problems in this
direction. Particularly, we consider the possibility for atom
laser beam to carry the {\it predesigned} angular momentum.

\par
The paper is organized as follows. In section 2 I consider the
quasi-hydrodynamic approach to a trapped Bose-gas below the BEC
transition temperature. Section 3 is devoted to the description of
the method for construction of Lagrange invariants and freezing-in
fields by means of the gauge transformation of the BEC
quasi-hydrodynamic equations. The applications of such invariants
to the creation of atom laser beam with predesigned parameters are
considered in Section 4.

\section{Quasi-hydrodynamic equations for the condensate atoms}

The starting point in description of BEC is the usual Heisenberg
equation of motion for the quantum field operator $\hat\psi ({\bf
r},t)$
 $${
 i\hbar\frac{\partial\hat\psi ({\bf r},t)}{\partial t} =
 \left[ \hat\psi ({\bf r},t), \hat{\mathcal H}\right] = \left\{
 -\frac{\hbar^2}{2m}\Delta+U^{(trap)} ({\bf r})\right\} \hat\psi
 ({\bf r},t)+g\vert\hat\psi ({\bf r},t)\vert^2\hat\psi ({\bf r},t)
 },\eqno(1)
 $$
where $U^{(trap)}({\bf r})$ is the confining potential, the
explicit form of which is not essential for us here. In equation
(1) we assumed $s$-wave short-range interatomic interaction with a
strength $g=4\pi ah^2/m$ ($a$ is the effective scattering length).
As usually, set
 $${
 \hat\psi({\bf r},t)=\Phi({\bf r},t)+\tilde\psi({\bf r},t)
 },\eqno(2)
 $$
where $\Phi({\bf r},t)=\langle\hat\psi({\bf r},t)\rangle$,
$\tilde\psi({\bf r},t)$ is the non-condensate field operator.
Taking an average of equation (1) with respect to a broken
symmetry nonequilibrium ensemble, we come to the equation for the
condensate wave function \cite{10}
 $${
 i\hbar\frac{d\Phi({\bf r},t)}{dt}=
 \left\{ -\frac{\hbar^2}{2m}\Delta+U^{(trap)}({\bf r})
 +gn_c({\bf r},t)+ 2g\tilde n({\bf r},t)
 \right\}\Phi({\bf r},t)+
 }
 $$

 $${
 g\tilde m({\bf r},t)\Phi^*({\bf r},t)
 +g\langle\tilde\psi^+({\bf r},t)\tilde\psi({\bf
 r},t)\tilde\psi({\bf r},t) \rangle
 },\eqno(3)
 $$
where $n_c({\bf r},t)=|\Phi({\bf r},t)|^2$ is the local density of
atoms in the condensate, 
$${ \tilde n({\bf r},t)= \langle
\tilde\psi^+({\bf r},t)\tilde\psi({\bf r},t)\rangle 
},\eqno(4) 
$$
is the non-equilibrium non-condensate density. Equation (4)
involves also the anomal non-condensate density $\tilde m({\bf
r},t)= \langle \tilde\psi({\bf r},t)\tilde\psi ({\bf r},t)\rangle$
and the three-field correlation function $\langle |\tilde\psi({\bf
r},t)|^2\tilde\psi({\bf r},t)\rangle$.  The appearance of two last
terms in (3) is a consequence of Bose broken symmetry in the
system.

\par
The earlier approaches to the equation (3) were based on the
assumption that all atoms are in the condensate. In this case the
so-called Gross--Pitaevskii \cite{17} equation appears: $${
i\hbar\frac{d\Phi({\bf r},t)}{dt}=
\left\{-\frac{\hbar^2}{2m}\Delta+ U^{(trap)}({\bf r})+gn_c({\bf
r},t) \right\}\Phi({\bf r},t) },\eqno(5) $$ which can be
conveniently rewritten in terms of the local condensate density
$${ n_c({\bf r},t)=|\Phi({\bf r},t)|^2 }\eqno(6a) $$ and local
velocity $${ {\bf v}_c({\bf r},t)=\frac{\hbar}{m}\nabla\theta({\bf
r},t) }.\eqno(6b) $$

Here $\theta({\bf r},t)$ is the phase of the condensate wave
function $${ \Phi({\bf r},t)=\sqrt{n_c({\bf
r},t)}\exp\left(i\theta({\bf r},t)\right) }.\eqno(7) $$

\par
It should be noted that the analogous nonlinear Shr\"odinger
equation can be obtained quite rigorously in the case of high
density, but weak enough point interaction \cite{18}. However, in
that case the physical sense of $\Phi$ is not clear enough.  So,
using (6) and (7), we can present equation (5) as the set of two
following equations $${ \frac{\partial n_c}{\partial
t}+\nabla(n_c{\bf v_c})=0 },\eqno(8a) $$

$${ \left\{ \frac{\partial}{\partial t}+ ({\bf v_c\nabla})
\right\}{\bf v_c}= -\frac{\nabla\mu_0}{m} },\eqno(8b) $$ where $${
\mu_0=-\frac{\hbar^2}{2m}\frac{\Delta\sqrt{n_c}}{\sqrt{n_c}} +
U^{(trap)}({\bf r})+gn_c({\bf r},t) }. $$ It is remarkable that
the equations (8) are hydrodynamic looking. This fact somewhat
confusing on the face of it. Really, equations similar to (8) was
obtained phenomenologycally to describe a superfluid component of
liquid helium \cite{19,20} which is a {\it strongly
interacting} many-particle system. Nevertheless we should keep in
mind that in Bose-Einstein condensed state we deal with the single
condensate wave function $\Phi ({\bf r}, t)$, which allows a
strong analogy with the order parameter in
superfluids.\footnote{Note, there is no complete microscopic
theory of superfluids until now. We only know that the dynamics of
superfluid component in superfluids can be well described by the
hydrodynamic equations similar to (8).}

The next step is to extend the preceding analysis to finite
temperatures where there is a large fraction of atoms outside of
the condensate. In this case we need two equations to be used.
While the condensate wave function can be described as earlier  by
the Gross--Pitaevskii equation, the distribution function of the
non-condensate atoms obeys a kinetic equation, which must take
into account the collisions of both types: non-condensate atoms
with each other and their interaction with a condensate.  The
quite rigorous derivation of the corresponding collision integrals
can be found in Appendix A of paper \cite{10}. After some
mathematics we come to the corresponding hydrodynamic--like
equations

$${ \frac{\partial n_c}{\partial t}+\nabla(n_c{\bf v}_c)=\Gamma
({\bf r},t) },\eqno(9a) $$

$${ \left\{ \frac{\partial}{\partial t}+ ({\bf v}_c\nabla)
\right\}{\bf v_c}= -\frac{\nabla\mu}{m} },\eqno(9b) $$ where $${
\Gamma ({\bf r},t)=-\int\frac{d^3p}{(2\pi\hbar)^3}J\left[ f({\bf
p},{\bf r},t)\right] },\eqno(10) $$ $J\left[ f({\bf p},{\bf
r},t)\right]$ is the collision integral corresponding to the
collisions between condensate and non-condensate atoms, which
functionally depends on the distribution function $f({\bf p},{\bf
r},t)$ of excited atoms. So, function $\Gamma ({\bf r},t)$ is the
characteristic rate of the atoms exchange between condensate and
non-condensate. New chemical potential $\mu$ in Eq. (9b) is now
defined by the relation $${
\mu=-\frac{\hbar^2}{2m}\frac{\Delta\sqrt{n_c}}{\sqrt{n_c}} +
U^{(trap)}({\bf r})+gn_c({\bf r},t)+2g\tilde{n}({\bf r},t)
},\eqno(11) $$ where $${ \tilde{n}({\bf
r},t)=-\int\frac{d^3p}{(2\pi\hbar)^3} f({\bf p},{\bf r},t) } $$ is
the density of non-condensate atoms. Equations (9) describe the
dynamics of BEC like an `effective fluid' with varying density.
Term $\Gamma ({\bf r},t)$ play the role of inhomogeneous and
nonstationary source.

\section{Hydrodynamic invariants}

As we realized in the previous section the evolution of BEC in the
frame of reasonable approximations can be described by the
equations, which are similar to the hydrodynamic ones. Classical
equations of ideal liquid have quite a number of invariants.
Except of ordinary integral invariants there are local invariants
as well. The Lagrange invariants and freezing-in invariants are
most important ones. Lagrange invariants are conserved along the
`liquid particles' trajectories, while the freezing-in invariants
are used in reference to the fields frozen into a liquid, i.e. the
corresponding physical quantity (field) vanishes in a frame which
moves with the fluid. In the papers \cite{21,22,23} a wide
class of invariants was found. Furthermore, the authors of Refs.
\cite{21,22} proposed the method of obtaining new invariants
on the basis of already known. The recent paper \cite{24} was
devoted to construction of the invariants of superfluid
hydrodynamic equations by means of their gauge transformation
\cite{25}. This method is very attractive because after the
gauge transformation the presence of many additional invariants
becomes obvious.

\par
The idea of gauge transformation \cite{24,25} can be modified
to be helpful in our case, i.e. BEC dynamics. Really, the local
condensate velocity is defined by the equation (6). If we wish the
condensate wave function to be single-valued, then the bypassing
along vortex line must lead to the change of a phase by the
integer of $2\pi$. To be so, we should do a cut. If the leap of
phase on the bank of cut is proportional to some new function, say
${\bf u}_c$, then we can do the following gauge transformation $${
{\bf v}_c=-\nabla\theta+\nabla\alpha+{\bf u}_c },\eqno(12) $$
where $\alpha$ is some gauge function. Gauge of the fields should
be done by the equation for $\nabla\theta-\nabla\alpha$ and by the
initial conditions. After the substitution of equation (12) into
(9b) we get $${ \left\{\frac{\partial}{\partial t}+({\bf
v}_c\nabla)\right\}u_{ci}= -\frac{\partial}{\partial
x_i}\left\{\mu+\frac{\partial}{\partial t} (\alpha -\theta
)\right\}-({\bf v}_c\nabla ) \left[\frac{\partial\alpha}{\partial
x_i}- \frac{\partial\theta}{\partial x_i} \right]- } $$

$${ -u_{cj}\frac{\partial v_{cj}}{\partial x_i}+ \left\{
v_{cj}+\frac{\partial}{\partial x_j}(\theta -\alpha)\right\}
\frac{\partial v_{cj}}{\partial x_i} }.\eqno(13) $$ It can be
easily tested that if the gauge function obeys the following
equation: $${ \left\{\frac{\partial}{\partial t}+({\bf
v}_c\nabla)\right\} (\theta -\alpha)=\mu-\frac{1}{2}v_c^2
},\eqno(14) $$ then equation (13) becomes $${
\left\{\frac{\partial}{\partial t}+({\bf v}_c\nabla)\right\}u_{ci}
=-u_{cj}\frac{\partial v_{cj}}{\partial x_i} }.\eqno(15) $$ The
gauge of field ${\bf u}_c$ is determined by equation (14) and by
the initial condition for $\theta$ or ${\bf u}_c$. The scalar
product of the field ${\bf u}_c$ and the flux line element
$\delta{\bf l}$ behaves like a mass element, which is conserved
along any trajectory \cite{24}. Direct test show that $${
\left\{\frac{\partial}{\partial t}+({\bf v}_c\nabla)\right\} ({\bf
u}_c\delta{\bf l})=0 }.\eqno(16) $$

\par
So, the quantity ${\bf u}_c\delta{\bf l}$ is the Lagrange
invariant. It should be noted that in this case the vorticity
${\bf w}=\mbox{rot}{\bf v}_c$ becomes freezing-in field
\cite{26}. Using the analogy with superfluid hydrodynamics we
believe that in BEC $\mbox{rot}{\bf v}_c=0$ everywhere except the
axes of vortices. So, we come to the following conclusion: if
there were vortices in the BEC initially and the invariant ${\bf
u}_c\delta{\bf l}$ is conserved, then there will be the given
conserved vorticity in any frame moving with the flux lines in
future.

\par
The possible existence of vortices in BEC has been under extensive
discussion for a rather long time (see {\it e.g.}
\cite{28,29,30,31,32,33,34}). And finally they were recently
obtained in the experiments \cite{35,36,37,38,39}.

In the next section we consider the consequences of this
conclusion for possible experimental realization of an atom laser.

\section{The stability of atom laser beam}
A number of atom laser schemes have been proposed during last few
years. Evident progress is already achieved in the realization of
pulsed lasers using a matter-wave splitter based on radio
frequency (rf) transitions \cite{15,40} and optical Raman
transitions \cite{41,42}. Such schemes, however, have several
shortcomings, the main of which is the difficulties in achieving a
continuous refilling. Another schemes, which allow to create the
continuous wave atom laser can be clearly divided into two
distinct classes: optical cooling \cite{43,44,45} and
evaporative cooling \cite{46,47,48,49}. Both models are
based in principle on the same idea: some source supplies atoms to
an upper-lying mode of an atom trap. This source mode is coupled
to the ground state mode (condensate) via a particular cooling
mechanism. It is hoped that the macroscopic population in this
ground state mode, or laser mode, can be built up and coupled to
outside world to produce the laser output. Independently on
particular model, cooling process or, in other words, process of
increase of the atom population in the ground state mode must
satisfy the main criterion: the uncontrolled perturbation of
ground state atoms should be minimal.

\par
I shall not keep myself in the frame of the particular
experiments (even successful, such as \cite{41,42} or
\cite{15,16}).
Below both the possible situations are considered.

\subsection{Continuous models}

In the previous section we found that the BEC can have Lagrange
invariants. Let us assume that initially all atoms in the trap are
in condensate. So, equation (9a) contains the quantity
$\Gamma({\bf r},t)=\nu_{p}({\bf r},t)$, which is just the rate of
pumping, i.e. the rate of increase of the population in ground
state due to the cooling of atoms from upper-lying mode. Let us
find the conditions for $\nu_{p}({\bf r},t)$, under which the
pumping of the ground state mode does not break the flux lines
(i.e. the all local invariants remain to be conserved). As it
follows from the equation (15), ${\bf u}_c\delta{\bf l}$ is always
a Lagrange invariant (see equation (16)) if the gauge condition
(14) is satisfied. However, if the flux lines are broken so that a
vorticity changes, then the condition (14) is necessarily broken
as well as ${\bf u}_c\delta{\bf l}$ becomes non-invariant. So, we
come to the simple conclusion: {\it to conserve a given vorticity
we must keep the regime of pumping to be such one to do the
condition (14) being always valid}. Obviously, the most simple
requirement is $\theta=\alpha$, or as it follows from equation
(14),
 $${
 \mu({\bf r},t)=\frac{1}{2}v_c^2({\bf r},t)
 }.\eqno(17)
 $$
Furthermore, this requirement automatically means that ${\bf
u}_c={\bf v}_c$ and the invariant ${\bf v}_v\delta{\bf l}$
contains the velocity itself.

\par
For simplicity consider the situation, when the velocity of laser
mode changes only in given direction. In this case, using
equations (9a), (11), and (17), we obtain
 $${
 \frac{\partial n_c}{\partial t}+\nu_p ({\bf r},t)+\nabla
 \left\{ \sqrt{2}n_c{\bf e}_v\sqrt{U^{(trap)}
 ({\bf r})+g(n_c+2\tilde{n})-\frac{\hbar^2}{2m}
 \frac{\Delta\sqrt{n_c}}{\sqrt{n_c}}} \right\}=0
 },\eqno(18)
 $$
where ${\bf e}_v={\bf v}_c/|{\bf v}_c|$. Result (18) gives the
connection between experimentally controllable quantities $n_c$,
$\tilde{n}$, $\nu_p ({\bf r},t)$, and $U^{(trap)}({\bf r})$. Of
course, this equation should further be solved numerically for
particular experimental configurations. Note that, of course the
relation (18) does not solve all problems of atom laser beam
stability, but it gives very useful tool for doing a choice of the
parameters of experimental setup. If the condition (18) is
satisfied, then at least the problem of angular momentum transfer
is solved in the frame of the done approximations.

\subsection{Pulsed models}

Here we consider the models of pulsed (not continuously refilled)
atom laser on the example of rf-transition scheme
\cite{15,16,40}. In this scheme the output coupler
includes resonant monochromatic radio frequency field transferring
atoms in some hyperfine state $F$ from the trapped into untrapped
magnetic sublevels. In the case of $^{23}$Na atoms F=1, so that
$s=-1$ corresponds to the trapped state, $s=0$ and $s=1$
corresponds to the untrapped and the repelled sublevels,
respectively (here $F$ is the total angular momentum, $s$ is the
magnetic quantum number). Equation (5) now becomes \cite{50}
 $${
 i\hbar\frac{d\tilde{\Phi}_s({\bf r},t)}{dt}=
 \left\{-\frac{\hbar^2}{2m}\Delta+ \hbar s\omega_{rf}
 U_s^{(trap)}({\bf r})+gn_s({\bf r},t) \right\}\tilde{\Phi}_s({\bf
 r},t)
 }
 $$

 $${ +\hbar\Omega\sum_{s'}(\delta_{s,s'+1}+\delta_{s,s'-1})
 \tilde{\Phi}_{s'}({\bf r},t)
 },\eqno(19)
 $$
where in rotating wave approximation
 $${
 \tilde{\Phi}_s({\bf r},t)=\mbox{e}^{-is\omega_{rf}t}
 \langle\hat{\psi}_s({\bf r},t)\rangle, \qquad s,s'\in\{-1,0,+1\}
 },\eqno(20)
 $$
$\hat{\psi}_s({\bf r},t)$ is the quantum field operator for atoms
belong the sublevel with given $s$, $\omega_{rf}$ is the frequency
of applied resonant rf field, $n_s=|\tilde{\Phi_s}|^2$. The
coupling constant
 $${
 \hbar\Omega=g\mu_{Bohr}\frac{|B|}{\sqrt{2}}
 }\eqno(21)
 $$
refers to the Rabi frequency due to the rf field. For a small
coupling strength the process of atoms leaking out of the
resonance points is faster than the Rabi oscillations. So we
further neglect the coupling into state $s=+1$ since it is
proportional to $\Omega^4$. Using such approximation and writing
the relation (7) for each sublevel $s$ we get for the density of
atoms in a laser beam $n_0$ the following relation, analogous to
the formula (9a):
 $${
 \frac{\partial n_0}{\partial t}+\nabla
 (n_0{\bf v}_0)= 2\Omega\sqrt{n_cn_0}\sin (\theta_c-\theta_0)
 },\eqno(22)
 $$
where $n_c\equiv n_{-1}$. Equation (22) has a clear physical
sense: variation of the atom beam density oscillates due to the
differences of the condensate and beam phases.

\par
Equation similar to (9b) looks
 $${
 \left\{
 \frac{\partial}{\partial t}+ ({\bf v}_0\nabla) \right\}{\bf v_0}=
 -\frac{\nabla\tilde{\mu}_0}{m}
 },\eqno(23)
 $$
where
 $${
 \tilde{\mu}_0=-\frac{\hbar^2}{2m}\frac{\Delta\sqrt{n_0}}{\sqrt{n_0}}
 + U_0^{(trap)}({\bf r})+gn_0({\bf r},t)+
 \hbar\Omega\sqrt{\frac{n_c}{n_0}} \cos (\theta_c-\theta_0)
 }\eqno(24)
 $$
is the new chemical potential. From equations (17), (22), (23),
and (24) we easily obtain the condition similar to (18):
 $${
 \frac{\partial n_0}{\partial t}+\nabla\left\{\sqrt{2}n_0{\bf e}_v
 \sqrt{U_0^{(trap)}({\bf r})+gn_0({\bf r},t)-\frac{\hbar^2}{2m}
 \frac{\nabla\sqrt{n_0}}{\sqrt{n_0}}+\hbar\Omega
 \sqrt{\frac{n_c}{n_0}}\cos (\theta_c-\theta_0)}\right\}
 }
 $$

 $${
 =2\Omega\sqrt{n_0n_c}\sin (\theta_c-\theta_0)
 }.\eqno(25)
 $$
Note, as it can be seen from the condition (25) the temporal
change of the beam atoms population depends on the phase
difference $(\theta_c-\theta_0)$, which is determined by the
frequency $\omega_{rf}$.

\par
Direct comparison of the formulae (18) and (25) shows that the
outcome (i.e. laser beam itself) in the continuous case can be
stabilized easier than in the pulsed regime.

\par
In conclusion, using the quasi-hydrodynamic approximations we have
found the local invariants of Bose-Einstein condensate in trapped
alkali gases. Particularly we obtained the Lagrange invariant,
which ensures the vorticity to be a freezing-in field. The
obtained results can be directly applied to the creation of highly
coherent atomic beam generators (atom lasers) with well controlled
angular momentum. Both the pulsed laser and laser with continuous
refilling are considered. The optimal conditions for the pumping
modes have been found (again for the both schemes).

\end{document}